\input harvmac

\def\Title#1#2{\rightline{#1}\ifx\answ\bigans\nopagenumbers\pageno0\vskip1in
\else\pageno1\vskip.8in\fi \centerline{\titlefont #2}\vskip .5in}
%

%
%
\ifx\includegraphics\UnDeFiNeD\message{(NO graphicx.tex, FIGURES WILL BE IGNORED)}
\def\figin#1{\vskip2in}
\else\message{(FIGURES WILL BE INCLUDED)}\def\figin#1{#1}
\fi
\def\Fig#1{Fig.~\the\figno\xdef#1{Fig.~\the\figno}\global\advance\figno
 by1}
%
%
%
%

%
%
\font\ticp=cmcsc10

\def\rightnouti{ {\alpha_{n}^{\rm{out} \ i }} }

\def\leftnouti{ {\tilde{\alpha}_{n}^{\rm{out} \ i }} }

\def\rightmoutj{ {\alpha_{m}^{\rm{out} \ j}} }

\def\leftmoutj{ {\tilde{\alpha}_{m}^{\rm{out} \ j}} }
\def\rightnini{ {\alpha_{n}^{\rm{in} \ i }} }

\def\leftnini{ {\tilde{\alpha}_{n}^{\rm{in} \ i }} }

\def\rightninj{ {\alpha_{n}^{\rm{in} \ j }} }
\def\rightminj{ {\alpha_{m}^{\rm{in} \ j}} }
\def\leftninj{ {\tilde{\alpha}_{n}^{\rm{in} \ j }} }
\def\leftminj{ {\tilde{\alpha}_{m}^{\rm{in} \ j}} }

\def\rightninjneg{ {\alpha_{-n}^{\rm{in} \ j }} }

\def\leftninjneg{ {\tilde{\alpha}_{-n}^{\rm{in} \ j }} }

\def\invac{ { \langle 0 _{\rm{in}}| } }
\def\vacin{ {|0_{\rm{in}} \rangle} }

\def\Xhat{{\hat X}}
\def\hf{{1\over 2}}
\def\e{{\epsilon}}
%
%
\lref\ACVone{
  D.~Amati, M.~Ciafaloni and G.~Veneziano,
  ``Superstring Collisions at Planckian Energies,''
  Phys.\ Lett.\  B {\bf 197}, 81 (1987).
}
\lref\ACVtwo{
  D.~Amati, M.~Ciafaloni and G.~Veneziano,
  ``Classical and Quantum Gravity Effects from Planckian Energy Superstring
  Collisions,''
  Int.\ J.\ Mod.\ Phys.\  A {\bf 3}, 1615 (1988).
}
\lref\GrMeone{
  D.~J.~Gross and P.~F.~Mende,
  ``The High-Energy Behavior of String Scattering Amplitudes,''
  Phys.\ Lett.\  B {\bf 197}, 129 (1987).
}
\lref\GrMetwo{
  D.~J.~Gross and P.~F.~Mende,
  ``String Theory Beyond the Planck Scale,''
  Nucl.\ Phys.\  B {\bf 303}, 407 (1988).
}
\lref\LQGST{
  S.~B.~Giddings,
  ``Locality in quantum gravity and string theory,''
  Phys.\ Rev.\  D {\bf 74}, 106006 (2006)
  [arXiv:hep-th/0604072].
}
\lref\AmatiXE{
  D.~Amati, M.~Ciafaloni and G.~Veneziano,
  ``Higher Order Gravitational Deflection And Soft Bremsstrahlung In Planckian
  Energy Superstring Collisions,''
  Nucl.\ Phys.\  B {\bf 347}, 550 (1990).
}

\lref\AiSe{
  P.~C.~Aichelburg and R.~U.~Sexl,
  ``On the Gravitational field of a massless particle,''
  Gen.\ Rel.\ Grav.\  {\bf 2}, 303 (1971).
}
\lref\EaGi{
  D.~M.~Eardley and S.~B.~Giddings,
  ``Classical black hole production in high-energy collisions,''
  Phys.\ Rev.\  D {\bf 66}, 044011 (2002)
  [arXiv:gr-qc/0201034].
}
\lref\tHoo{
  G.~'t Hooft,
  ``Dimensional reduction in quantum gravity,''
  arXiv:gr-qc/9310026.
  }
\lref\Suss{
  L.~Susskind,
  ``The World as a hologram,''
  J.\ Math.\ Phys.\  {\bf 36}, 6377 (1995)
  [arXiv:hep-th/9409089].
}
\lref\Bouss{
  R.~Bousso,
  ``The holographic principle,''
  Rev.\ Mod.\ Phys.\  {\bf 74}, 825 (2002)
  [arXiv:hep-th/0203101].
}
\lref\GiTr{S.~B.~Giddings,
  ``Quantum mechanics of black holes,''
  arXiv:hep-th/9412138\semi
  ``The black hole information paradox,''
  arXiv:hep-th/9508151.
}
\lref\Astrorev{
  A.~Strominger,
  ``Les Houches lectures on black holes,''
  arXiv:hep-th/9501071.
}
\lref\Mald{
  J.~M.~Maldacena,
  ``The large N limit of superconformal field theories and supergravity,''
  Adv.\ Theor.\ Math.\ Phys.\  {\bf 2}, 231 (1998)
  [Int.\ J.\ Theor.\ Phys.\  {\bf 38}, 1113 (1999)]
  [arXiv:hep-th/9711200].
}
\lref\Vene{
  G.~Veneziano,
  ``A Stringy Nature Needs Just Two Constants,''
  Europhys.\ Lett.\  {\bf 2}, 199 (1986); talk presented at the Italian Physical Society Meeting, Naples, 1987.
}
\lref\Gross{D.~J.~Gross,
  ``Superstrings And Unification,'' PUPT-1108
{\it Plenary Session talk given at 24th Int. Conf. on High Energy Physics, Munich, West Germany, Aug 4-10, 1988}.
}
\lref\GMH{
  S.~B.~Giddings, D.~Marolf and J.~B.~Hartle,
  ``Observables in effective gravity,''
  Phys.\ Rev.\  D {\bf 74}, 064018 (2006)
  [arXiv:hep-th/0512200].
}
\lref\GaGi{
  M.~Gary and S.~B.~Giddings,
  ``Relational observables in 2d quantum gravity,''
  arXiv:hep-th/0612191, Phys.\ Rev.\  D {\bf 75} 104007 (2007).
}
\lref\LPSTU{
  D.~A.~Lowe, J.~Polchinski, L.~Susskind, L.~Thorlacius and J.~Uglum,
  ``Black hole complementarity versus locality,''
  Phys.\ Rev.\  D {\bf 52}, 6997 (1995)
  [arXiv:hep-th/9506138].
}
\lref\YoNa{
  H.~Yoshino and Y.~Nambu,
  ``Black hole formation in the grazing collision of high-energy particles,''
  Phys.\ Rev.\  D {\bf 67}, 024009 (2003)
  [arXiv:gr-qc/0209003].
}
\lref\GiRy{
  S.~B.~Giddings and V.~S.~Rychkov,
  ``Black holes from colliding wavepackets,''
  Phys.\ Rev.\  D {\bf 70}, 104026 (2004)
  [arXiv:hep-th/0409131].
}
\lref\DvSa{
  H.~J.~de Vega and N.~G.~Sanchez,
  ``Quantum string scattering in the Aichelburg-Sexl geometry,''
  Nucl.\ Phys.\  B {\bf 317}, 706 (1989).
}
\lref\HoSt{
  G.~T.~Horowitz and A.~R.~Steif,
  ``Strings in strong gravitational fields,''
  Phys.\ Rev.\  D {\bf 42}, 1950 (1990).
}
\lref\GSW{
  M.~B.~Green, J.~H.~Schwarz and E.~Witten,
 {\sl Superstring theory. vol. 1: Introduction,}
 Cambridge, UK: Univ. Pr. (1987).
}
\lref\Froiss{
  M.~Froissart,
  ``Asymptotic behavior and subtractions in the Mandelstam representation,''
  Phys.\ Rev.\  {\bf 123}, 1053 (1961).
}
\lref\CeMa{F. Cerulus and A. Martin, ``A lower bound for large angle elastic 
scattering at high energies,''  Phys. Lett. {\bf 8}, 80 (1964).}
\lref\Mart{A. Martin, ``Minimal interactions at very high transfers,''
 Nuov. Cim. {\bf 37}, 671 (1965).}
\lref\Penunpub{R.Penrose
{\sl unpublished} (1974).}
\lref\KKS{
  M.~Karliner, I.~R.~Klebanov and L.~Susskind,
  ``Size And Shape Of Strings,''
  Int.\ J.\ Mod.\ Phys.\  A {\bf 3}, 1981 (1988).
}
\lref\Hsu{
  S.~D.~H.~Hsu,
  ``Quantum production of black holes,''
  Phys.\ Lett.\  B {\bf 555}, 92 (2003)
  [arXiv:hep-ph/0203154].
}
\lref\FPVV{
  M.~Fabbrichesi, R.~Pettorino, G.~Veneziano and G.~A.~Vilkovisky,
  ``Planckian energy scattering and surface terms in the gravitational
 action,''
  Nucl.\ Phys.\  B {\bf 419}, 147 (1994).
}
\lref\VenezianoDF{
  G.~Veneziano,
  ``Mutual Focusing of Graviton Beams,''
  Mod.\ Phys.\ Lett.\  A {\bf 2}, 899 (1987).
}
\lref\GiddingsBE{
  S.~B.~Giddings,
 ``(Non)perturbative gravity, nonlocality, and nice slices,''
  Phys.\ Rev.\  D {\bf 74}, 106009 (2006)
  [arXiv:hep-th/0606146].
}
\lref\BanksGD{
  T.~Banks and W.~Fischler,
  ``A model for high energy scattering in quantum gravity,''
  arXiv:hep-th/9906038.
}
\lref\VenezianoER{
  G.~Veneziano,
  ``String-theoretic unitary S-matrix at the threshold of black-hole
  production,''
  JHEP {\bf 0411}, 001 (2004)
  [arXiv:hep-th/0410166].
}
\lref\ArkaniHamedKY{
  N.~Arkani-Hamed, S.~Dubovsky, A.~Nicolis, E.~Trincherini and G.~Villadoro,
  ``A Measure of de Sitter Entropy and Eternal Inflation,''
  arXiv:0704.1814 [hep-th].
}
\lref\AmatiXE{
  D.~Amati, M.~Ciafaloni and G.~Veneziano,
  ``Higher Order Gravitational Deflection And Soft Bremsstrahlung In Planckian
  Energy Superstring Collisions,''
  Nucl.\ Phys.\  B {\bf 347}, 550 (1990).
}
\lref\AmatiTB{
  D.~Amati, M.~Ciafaloni and G.~Veneziano,
  ``Effective action and all order gravitational eikonal at Planckian
  energies,''
  Nucl.\ Phys.\  B {\bf 403}, 707 (1993).
}
\lref\MuSo{
  I.~J.~Muzinich and M.~Soldate,
  ``High-Energy Unitarity of Gravitation and Strings,''
  Phys.\ Rev.\  D {\bf 37}, 359 (1988).
}
\lref\ChFi{
  M.~Chaichian and J.~Fischer,
  ``Higher dimensional space-time and unitarity bound on the scattering amplitude,"
  Nucl.\ Phys.\  B {\bf 303}, 557 (1988).
}
\lref\CFV{
  M.~Chaichian, J.~Fischer and Yu.~S.~Vernov,
  ``Generalization Of The Froissart-Martin Bounds To Scattering In A Space-Time
  Of General Dimension,''
  Nucl.\ Phys.\  B {\bf 383}, 151 (1992).
}
\lref\SusskindAA{
  L.~Susskind,
  ``Strings, black holes and Lorentz contraction,''
  Phys.\ Rev.\  D {\bf 49}, 6606 (1994)
  [arXiv:hep-th/9308139].
}
\lref\PetrovHB{
  V.~A.~Petrov,
  ``Froissart-Martin bound in spaces with compact extra dimensions,''
  Mod.\ Phys.\ Lett.\  A {\bf 16}, 151 (2001)
  [arXiv:hep-ph/0008329]\semi
  V.~A.~Petrov,
 ``Froissart-Martin bound in spaces of compact dimensions,''
  Phys.\ Atom.\ Nucl.\  {\bf 65}, 877 (2002)
  [Yad.\ Fiz.\  {\bf 65}, 909 (2002)].
  }
 \lref\KaNa{
  K.~Kang and H.~Nastase,
  ``Planckian scattering effects and black hole production in low M(Pl)
  scenarios,''
  Phys.\ Rev.\  D {\bf 71}, 124035 (2005)
  [arXiv:hep-th/0409099].
}
\Title{\vbox{\baselineskip12pt
\hbox{NSF-KITP-07-115}
}}
{\vbox{\centerline{Gravitational effects in}\centerline{ultrahigh-energy string scattering}
}}
\centerline{{\ticp Steven B. Giddings,${}^a$\footnote{$^\ast$}{Email address: giddings@physics.ucsb.edu} }  {\ticp David J. Gross,${}^{b}$\footnote{$^\ddagger$}{Email address: gross@kitp.ucsb.edu}} and {\ticp Anshuman Maharana}${}^a$\footnote{$^\dagger$}{Email address: anshuman@physics.ucsb.edu}
}
\centerline{${}^a$\sl Department of Physics}
\centerline{and}
\centerline{${}^b$\sl Kavli Institute of Theoretical Physics}
\centerline{\sl University of California}
\centerline{\sl Santa Barbara, CA 93106}
\bigskip
\centerline{\bf Abstract}

Ultrahigh-energy string scattering is investigated to clarify the relative role of string and gravitational effects, and their possible contributions to nonlocal behavior.  Different regimes can be characterized by varying the impact parameter at fixed energy.  In the regime where momentum transfers reach the string scale, string effects appear subdominant to higher-loop gravitational processes, approximated via the eikonal.  At smaller impact parameters, ``diffractive" or ``tidal" string excitation leads to processes dominated by highly excited strings.  However, new evidence is presented that these excitation effects do not play a direct role in black hole formation, which corresponds to breakdown of gravitational perturbation theory and appears to dominate at sufficiently small impact parameters.   The estimated  amplitudes violate expected bounds on high-energy behavior for local theories.

\Date{}

\newsec{Introduction}

High-energy scattering is a time-tested method to probe the fundamental dynamics of a theory.  For this reason, there has been significant effort and progress towards determining the high-energy scattering behavior of string theory  -- see for example \refs{\GrMeone\GrMetwo\ACVone\ACVtwo\AmatiXE\VenezianoER-\LQGST}.  

A particularly interesting issue relates to the status of locality in string theory.  There is a widespread feeling, and various evidence, that string theory, being a theory of extended objects, is indeed in some sense not as local as quantum field theory.  This certainly seems to be true at the string scale, but even more interesting is the question of whether string theory manifests longer-scale, and even macroscopic, nonlocalities.  Certainly such behavior could nicely accord with suggestions that string theory is {\it holographic}\refs{\tHoo\Suss-\Bouss}, that is, can be completely described by a number of degrees of freedom that grows with the surface area bounding a volume, or even is completely captured by a theory on such a boundary, as in AdS/CFT\refs{\Mald}.  Possible nonlocality is also of great interest for understanding whether and how string theory resolves a central problem in gravity, that of black hole information.\foot{For reviews see \refs{\Astrorev,\GiTr}.} 

Locality in a theory can typically be investigated either through behavior of local observables, or more indirectly through high-energy scattering.  In a gravitational theory such as string theory, precisely local observables should not exist, although there has been some success in understanding how approximately local observables emerge in appropriate limits from relational observables\refs{\GMH,\GaGi}.  It thus becomes of great interest to better understand high-energy string scattering.  While scattering at a threshhold energy near the string scale should exhibit a variety of novel phenomena such as excited string states, as well as string-scale nonlocalities\refs{\GrMeone,\GrMetwo},
possibly even more profound is the ultrahigh-energy regime far beyond the string scale.  For example, at ultrahigh energies, one has sufficient energy to excite a macroscopic string, of length $L\sim E/M_s^2$.  These observations have motivated proposal of a string uncertainty principle\refs{\Vene,\Gross}, and were suggested to play a role in resolving the information paradox\refs{\LPSTU}.  It is important to determine whether such macroscopic  nonlocality  due to string extendedness is indeed a feature of the theory.

Despite the progress in understanding high-energy string scattering\refs{\GrMeone\GrMetwo\ACVone\ACVtwo\AmatiXE-\VenezianoER}, various puzzles have remained\refs{\LQGST}, and a clear and complete picture has been lacking.  Particularly important is understanding the relative role of effects due to string excitation, such as long strings, and effects that are more easily described as being essentially gravitational.  
For example:  if excitation of strings of length $\propto E$ is an important effect in an ultrahigh-energy collision, this would seem to interfere with the very formation of a black hole, since it could prevent sufficient concentration of energy to form a trapped surface.

Indeed, one might expect string excitation to become important for momentum transfers of order $M_s$, and in particular, refs.~\refs{\ACVone\ACVtwo-\AmatiXE} have argued that string excitation becomes important at impact parameters far greater than those required to form a black hole.  These observations raise both the prospect of some essentially stringy long-range nonlocal effects, as well as the possibility that black holes don't form in high-energy string collisions.

This paper will investigate these questions more closely.  We begin in section two with a summary of some of the basic effects in string scattering, at ultrahigh energy, and as we vary the impact parameter to control the strength of the interaction.  Section three then explores the role of string excitation in scattering at momenta transfers $\sim M_s$, and provides arguments that in this regime higher-loop gravitational amplitudes dominate over string effects.  Section four then proceeds to smaller impact parameters; here ``diffractive string excitation\refs{\ACVtwo}," which can be understood as arising from gravitational tidal distortion\LQGST, becomes an important effect, and ultimately overwhelms purely elastic scattering.  Nonetheless, following more heuristic arguments presented in \LQGST, we provide calculations strongly supporting a picture in which black hole formation proceeds without interference from such tidal excitation.

Section five briefly summarizes the overall picture, and resulting puzzles.  In short, we find {\it no} evidence that extendedness of the string is a mechanism producing long-distance nonlocal effects in high-energy scattering,  and in particular no evidence that such effects interfere with black hole formation.
In contrast, the black hole threshhold raises important questions of locality.  The origin of the threshold is breakdown of the perturbation series.  Precise prediction would appear to require a fully non-perturbative theory, string theory or otherwise; order-by-order UV finiteness of the perturbation series does not suffice to complete the theory in this regime.  Moreover, such a theory would bring us face to face with the information problem.  Rough features of the resulting amplitudes can be inferred from the semiclassical description of black holes, and violate expected general bounds on local field theories, of the Froissart and Cerulus-Martin form.   Both these facts, and the statement that a unitary resolution of the information problem apparently requires nonlocality on the scale of the black hole, are suggestive of nonlocal effects that are intrinsically gravitational in nature.

\newsec{Overview of scattering regimes}

We will consider scattering in string theory compactified to $D$ dimensions, in the ultrahigh energy limit $E\gg M_s$, $E\gg M_D$.  Here $M_s\sim1/\sqrt{\alpha'}$ is the string mass scale, and the  Planck mass $M_D$ is given in terms of the $D$-dimensional Newton's constant  by
\eqn\planckdef{M_D^{2-D}= G_D\propto g_s^2 M_s^{2-D}\ .}
Below the ultrahigh energy regime, the distinction between $M_s$ and $M_D$ plays an important role in the dynamics, in controlling the relative contributions of string and gravitational effects.  Our goal is to determine what effects dominate dynamics in the ultrahigh energy region.

A useful way to parametrize ones entry into this region is to describe scattering as a function both of the energy $E=\sqrt s$ and of the impact parameter $b$ of the collision -- the later is, after all, something over which we have a large degree of control  in everyday experiments.  Moreover, for sufficiently large $b$, interactions are very weak.  Thus, one can imagine gradually decreasing $b$, for fixed ultrahigh  $E$, and asking what dynamics comes into play as one does so.

At very large distances, exchange of massless fields dominates the dynamics, and since its effective coupling grows with $E$, gravity is dominant among these.  A central question is what other effects become important with decreasing $b$.  In particular, since strings have internal structure, one might expect important effects when this structure is excited.  
There is extensive literature on this subject (see for example \refs{\GrMeone\GrMetwo\ACVone\ACVtwo\AmatiXE\VenezianoER-\LQGST}), but also some outstanding puzzles.

Let us describe some of the known features, and puzzles.  First, one might na\"\i vely expect 
string behavior to become apparent at  impact parameters 
\eqn\longstr{b\sim M_s^{-2} E\ ,}
where there is sufficient energy to create a stretched string of length $b$.  However, there is no indication for such string effects  in scattering.  For example, the tree-level amplitude at these impact parameters is well-approximated by graviton exchange, and a good explanation for this is that long strings in the $s$-channel are dual to exchange of the graviton mode of the string in the $t$ channel.  

Indeed, the results of \refs{\ACVone,\ACVtwo} suggest that scattering is simply dominated by long-range gravity until one reaches a regime of  ``diffractive excitation."  A simple mechanical description of this phenomenon, as ``tidal excitation" of one string in the gravitational field of the other, was given in \LQGST.  This tidal string excitation becomes important at impact parameters
\eqn\diffst{b_D\sim {1\over M_D} \left({E\over M_s}\right)^{2\over D-2}\ .}

Ref.~\LQGST\ noted that as $b$ decreases beyond $b_D$, tidal excitation becomes sufficiently large that a string can become stretched to scales comparable to the impact parameter.  Such ``large tidal excitation" is expected to occur at impact parameters of order
\eqn\bT{b_T \sim \left({G_DE^2\over M_s^3}\right)^{1\over D-1} \ .}
Large tidal excitation raises the prospect of non-trivial string effects, and perhaps that of some sort of stringy nonlocality.  Moreover, the impact parameter \bT\ is {\it larger} than that for black hole formation, which is given by the Schwarzschild radius of the center-of-mass energy,
\eqn\rsch{b\sim R_S(E)\sim (G_D E)^{1\over D-3}\ .}
This means that such large tidal excitation raises the prospect that black holes wouldn't form in high-energy string collisions, because the string energy distribution is spread out on scales large as compared to the would-be horizon.  To see this, notice that \refs{\ACVtwo} shows that the ``elastic" part of the amplitude falls exponentially in $G_D s/M_s^2b^{D-2}$ as tidal excitation takes over.  Near the horizon radius, the amplitude for unexcited string scattering is thus exponentially small in
\eqn\deltael{\delta_{\rm el}(R_S)\sim  E^{D-4\over D-3}/G_D^{1\over D-3} M_s^2\ .}

In fact, even {\it before} reaching these impact parameters, one is lead to consider other possibly important string effects.  Specifically, the asymptotic form of the four-graviton superstring amplitude, as $s\rightarrow\infty$ with fixed $t$, is\foot{See, for example, \refs{\GSW}.} (working in units $\alpha'=1/2$)
\eqn\fourgrav{{\cal A}^{\rm string}_{0}(s,t)\propto g_s^2  {\Gamma(-t/8)\over \Gamma(1+t/8)} s^{2+t/4} e^{2-t/4}\ .}
For $-t= q^2\sim 1$,  this amplitude shows significant corrections to the tree-level amplitude of pure gravity,
\eqn\gravexch{{\cal A}^{\rm grav}_{0}(s,t) \propto G_D {s^2\over t} \ ,}
due to string excitation.  
The condition $q^2\sim1$ corresponds to an impact parameter
\eqn\btone{b_t\sim \left(G_D E^2/M_s\right)^{1\over D-3}\ , }
and
raises the question of why the picture of eikonal scattering down to at least the impact parameter $b_D$ of tidal string excitation makes sense.

\newsec{Scattering at $q^2\sim 1$}

We turn first to the second question, justifying the neglect of string effects  when the impact parameter reaches the region \btone\ where $q^2\sim 1$.  Specifically, the eikonal approximation sums an infinite class of loop diagrams that contribute to the scattering.  The question at hand is the extent to which string corrections modify this sum.  To estimate the loop amplitudes in the ultrahigh-energy regime, one may follow the eikonal method, but capture string effects by sewing together string tree amplitudes \fourgrav\ rather than those for single graviton exchange. 
The $N$-loop term in this sum takes the form (see {\it e.g.} \refs{\MuSo})
\eqn\Nloop{i{\cal A}_N^{\rm string} =  {2s\over (N+1)!}\int \left[\prod_{j=1}^{N+1} {d^{D-2}k_j\over (2\pi)^{D-2} }{i{\cal A}_0^{\rm string}(s,-k_j^2)\over 2s} \right] (2\pi)^{D-2} \delta^{D-2}\left(\sum_j k_j - q_\perp\right) \ .}
Here in the high-energy limit, the sum over intermediate propagators approximately yields on-shell projectors, which in turn enforce the condition that the exchanged momenta  be perpendicular to the center-of-mass collision axis;  $q_\perp$ is the corresponding projection of the momentum transfer.

One can first ask, for a given order $N$, what momentum configuration dominates the integral.  Notice that as $q$ increases, the tree amplitude \fourgrav\ decreases.  In particular, string corrections lead to exponential weakening of the amplitude with increasing $q$.  This indicates that the ``optimal" distribution of momentum is such that the string corrections are minimized, with each momentum $k_j\sim q/N$.  As a result, the net correction is a factor of the form 
\eqn\corrfact{s^{-q^2/4N}}
and is comparatively small for $q^2\sim 1$ and $N\gg1$.  In short, the exponential softening due to string effects only reinforces the tendency of the momentum transfer to distribute uniformly over the exchanged strings.

We next examine which $N$ dominate the amplitude.  If we define the eikonal phase
\eqn\eikph{\chi(x) = {1\over 2s} \int {d^{D-2}k\over (2\pi)^{D-2}} e^{ik_\perp\cdot x} {\cal A}_0(s,t)\ ,}
the sum of the amplitudes \Nloop\ takes the eikonal form
\eqn\eikamp{i{\cal A}_{\rm eik}(s,t) = 2s \int d^{D-2} x_\perp e^{-iq_\perp x_\perp}  (e^{i\chi}-1)\ .}
For the pure gravity amplitude \gravexch, the eikonal phase takes the form
\eqn\eikphv{\chi(x) \sim G_D {s \over b^{D-4}}\ ,}
where $x_\perp = b$; for $D=4$ it is a logarithm.  The dominant loop order in the eikonal sum thus occurs near
\eqn\Ndom{N\sim {G_D E^2 \over b^{D-4}}\ .}
Thus, for the impact parameter \btone\ where $q^2\sim 1$, the dominant amplitudes are at very high loop order,
\eqn\Nt{N\sim (G_D E^2 M_s^{D-4})^{1\over D-3}\ .}

These arguments strongly suggest the consistency of replacing string amplitudes by pure gravity amplitudes in the regime $q^2\roughly <1$.  Note that, of course, the full $N$-loop gravity amplitudes will be UV divergent.  However, these UV divergences are not visible in the leading order expansion in $E$, which gives the eikonal amplitude  \eikamp.  The UV divergences are short-distance effects, and would not be expected to play a role in the physics at the large impact parameters we are considering.  This statement should be true for any consistent regulator of gravity, and in this vein we can regard string theory as providing such a regulator.  The above arguments suggest that amplitudes down to the regime $q^2\roughly< 1$ do not significantly depend on such a regulator -- long distance dynamics of gravity is dominant.

As expected, the eikonal amplitudes \eikamp\ closely correspond to a semiclassical picture.  Indeed, \refs{\ACVtwo} has argued that their form matches a picture of one string scattering in the classical metric of the other; for an ultra-high energy string  this is the Aichelburg-Sexl metric\refs{\AiSe}.  The combined classical metric of the pair of colliding strings is that of colliding Aichelburg-Sexl metrics, which  was shown in \refs{\Penunpub\EaGi-\YoNa} to form a closed trapped surface, and thus black hole, at impact parameters of order $R_S(E)$.  However, before concluding that black holes form in the collision of a pair of strings, one needs to check that other effects don't intervene as we decrease the impact parameter to this value.  The specific effects of concern are the tidal string excitation effects described above.

\newsec{Tidal string excitation and black hole formation}

\subsec{Collision with an Aichelburg-Sexl metric}

The preceding section argued that graviton exchange, in the eikonal approximation, is the dominant dynamics of scattering at least until the regime where tidal string excitation becomes large.  However, as we've said, once the expected tidal stretching of strings reaches a size comparable to the impact parameter, one should address the question\LQGST\ of whether the strings become so excited that this leads to new nonlocal effects.  If so, one might be concerned that there is no meaningful sense in which a black hole forms.  Ref.~\LQGST\ outlined arguments that black hole formation should in fact remain relevant, but these bear closer scrutiny.

As stated, the collision of two localized high-energy objects is classically well-described by the collision of two Aichelburg-Sexl\refs{\AiSe} metrics.  Penrose argued that the combined metric with zero impact parameter forms a trapped surface, and \refs{\EaGi} gave a construction\foot{For $D>4$ the construction was made explicit via numeric methods in \refs{\YoNa}.} of trapped surfaces for $b\roughly<R_S(E)$.  Thus, by the area theorem, a black hole forms.  Since the curvature at the trapped surface is small, modulo a mild singularity from the point particle limit (which is smoothed out for quantum wavepackets\refs{\GiRy}), one expects classical black hole formation to well-approximate the quantum process.\foot{For related discussion, see \refs{\BanksGD,\Hsu}.}

The specific concern with delocalization of strings is that if the strings are spread out on scales large compared to the horizon, there is no meaningful sense in which their energy is concentrated inside a region small enough to form a black hole.  A full treatment of this problem apparently requires a non-perturbative formulation of high-energy string scattering.  However, if there is such effect, it seems extremely probable that it would be seen in the approximation where one string scatters in the approximate Aichelburg-Sexl metric of the other string, and where backreaction is neglected.  Certainly in this approximation one expects the essential effect of  tidal string excitation and spreading -- the question is whether it is sufficient to prevent black hole formation.

Working in the center-of-mass frame with motion in the $z$ direction, define null coordinates 
\eqn\nullc{u=t-z\ ,\ v=t+z}
and transverse coordinates $x^i$, with transverse distance $\rho=\sqrt{x^ix^i}$.
The Aichelburg-Sexl metric for the right-moving string of energy $\mu=E/2$ is 
\eqn\as{ ds^{2} = -du dv + dx^{i} dx^{i} +\Phi(\rho) \delta (u) du^2}
with
\eqn\fd{\eqalign{ \Phi(\rho) &=  -8G_D\mu\ln\rho\quad ,\quad D=4\ ,\cr
&= {16\pi G_D\mu\over \Omega_{D-3} (D-4) \rho^{D-4}}\quad ,\quad D>4\ ,}}
and $\Omega_{D-3}$ the area of the unit $D-3$ sphere.

We will consider motion of the second, left-moving, string in this background, neglecting its backreaction on the full metric.  Quantization in such plane-fronted waves has been studied in \refs{\DvSa,\HoSt}.  We begin by reviewing and elaborating on their results.

\subsec{Light-cone quantization}

For the metric \as, the conformal-gauge sigma-model action is
\eqn\lcact{S= -{1\over 4\pi \alpha' } \int d\tau d\sigma \left[ -\partial_aU\partial^aV +\partial_aX^i \partial^a X^i + \Phi(X^i) \delta(U) \partial_a U\partial^a U\right]\ .}
The light-cone structure significantly aids in the quantization.  Specifically, define light cone gauge through the coordinate $u$:
\eqn\lcgauge{U(\tau) = 2\alpha' p^u \tau\ .}

The form of \fd\ shows that the action is non-linear and thus non-trivial.  However, one may work to leading order in the $\alpha'$ expansion.  Specifically, let $x^\mu(\tau)$ describe the classical trajectory of the center-of-mass of the string,  taken to be a null geodesic in the metric \as, and expand the string trajectory about this:
\eqn\strtraj{X^\mu(\tau,\sigma)=x^\mu(\tau) + {\hat X}^\mu(\tau,\sigma)\ .}
Correspondingly, $\Phi$ is expanded about the impact point of the null trajectory as
\eqn\phiexp{\Phi(X^i) = \Phi(x^i) + \Xhat^j(\tau,\sigma) \partial_j\Phi(x^i) + \hf \Xhat^j \Xhat^k \partial_j\partial_k \Phi(x^i)+\cdots\ .}
This is thus an expansion in $\Xhat\partial\sim \Xhat/b$.  
We will work up to quadratic order in this expansion.  A test of the validity of this approximation will be to compute a typical value of $\Xhat/b$ at this order.

Away from the shock front at $\tau=0$, the string propagates freely.  Thus in the ``in" and ``out" regions before and after the collision, we expand\foot{Our conventions for mode expansion of the the string coordinate
differs from that of \DvSa, we use the more standard practice of
including an explicit factor of $i$ multiplying the Fourier
coefficients.}
\eqn\modeexout{ X^{i} = x^{\rm{out}\ i}_{0} + 2\alpha'p^{\rm{out} \ i}\tau
 + i \sqrt{ \alpha' \over 2 } \sum_{ n
\neq 0} {1\over n} \left[ \alpha_{n}^{\rm{out} \ i} e^{-in(\tau+\sigma)} +  \tilde{\alpha}_{n}^{\rm{out} \ i} e^{-in(\tau-\sigma)}\right]\ ,\  \tau > 0 }
and
\eqn\modeexin{ X^{i} = x^{\rm{in}\ i}_{0} + 2\alpha'p^{\rm{in} \ i}\tau
 + i \sqrt{ \alpha' \over 2 } \sum_{ n
\neq 0} {1\over n} \left[ \alpha_{n}^{\rm{in} \ i} e^{-in(\tau+\sigma)} +  \tilde{\alpha}_{n}^{\rm{in} \ i} 
e^{-in(\tau-\sigma)}\right]\ ,\  \tau <0\ . }
The oscillators satisfy the usual commutation relationships
\eqn\com{[\rightnouti, \rightmoutj ] = [\leftnouti, \leftmoutj ] =
[\rightnini, \rightminj ] = [\leftnini, \leftminj ] = n
\delta_{m+n} \delta^{ij}}
\eqn\comt{ [\rightnouti,\leftmoutj] = [\rightnini,\leftminj] = 0 \ .}
For the remainder of the paper, we set $\alpha'=1/2$.

The relationship between the {\it in} and {\it out} oscillators
is determined by string propagation through the interface at
$\tau=0$.  The quantum equation of motion following from the action \lcact\ is
\eqn\qeom{(\partial_\tau^2 - \partial_\sigma^2) X^i - \hf p^u \partial^i \Phi \delta(\tau) =0\ .}
This should be supplemented by the constraints, $T_{++}=T_{--}=0$, which determine the solution for $X^-$ in terms of the $X^i$.  Matching the solutions \modeexout, \modeexin\ across the shock gives the conditions
\eqn\matcheq{\eqalign{ \left[\partial_\tau X^i_{\rm out} - \partial_\tau X^i_{\rm in} \right]\vert_{\tau=0} &= {p^u\over 2} \partial^i \Phi\cr  X^i_{\rm out}(0, \sigma) &= X^i_{\rm in}(0,\sigma)\ .}}
These conditions can be easily solved by writing
\eqn\lrdecomp{X^i_{\rm in,out} = X^{i\,R}_{\rm in,out}(\sigma-\tau) + X^{i\,L}_{\rm in,out}(\sigma+\tau)\ .}
Specifically, $X_{out}$ is determined in terms of $X_{in}$ by the expression
\eqn\inout{X^i_{\rm out}(\tau,\sigma) = X^i_{\rm in}(\tau,\sigma) + {p^u\over 4} \left[ \int_{\sigma-\tau}^{\sigma+\tau} d\sigma' \partial^i \Phi\left(X^j(0,\sigma')\right)\right]\ .}

To quadratic order in the expansion \phiexp, these equations are linear and can be readily solved to relate the oscillators  by a Bogoliubov transformation.  In particular, the individual {\it out} oscillators can be read off from the expansions of $X^{i\,L}_{out}$, $X^{i\,R}_{out}$.  The result is 
\eqn\btright{ \rightnouti = ( \delta^{i}_{j} + {ip^u\over 4n}\Phi_{ j}^{i})
\rightninj - {ip^u\over 4n}\Phi_{ j}^{i} \leftninjneg }
and
\eqn\btleft{ \leftnouti = ( \delta^{i}_{j} +{ip^u\over 4n}\Phi_{ j}^{i})
\leftninj - {ip^u\over 4n}\Phi_{ j}^{i} \rightninjneg }
where
\eqn\betadef{\Phi_{ij} = \partial_i\partial_j \Phi(x^k)\ .}

\subsec{String size}

To estimate the effects of string-spreading on black hole formation, we next compute the characteristic spread for a string initially in its ground state that propagates through the Aichelburg-Sexl metric.  We will do so by calculating the correlator
\eqn\deltadef{ \invac [ X^{i}_{\rm out} - x^{i}(\tau)] [ X^{j}_{\rm out}
- x^{j}(\tau)] \vacin =  \invac\Xhat_{\rm out}^i(\tau,\sigma) \Xhat_{\rm out}^j(\tau,\sigma) \vacin , \ \tau > 0 }
determining the distribution about the center of mass trajectory.
Inserting the expansion \modeexout\ and the relations \btright\ and \btleft, 
one finds after some simple algebra
\eqn\deltares{\invac\Xhat^i(\tau,\sigma) \Xhat^j(\tau,\sigma) \vacin =  \sum_{n=1}^\infty \left[{\delta^{ij}\over 2n}
+ {1\over4 n^2} p^u\Phi^{ij} \sin(2n\tau) +{1\over8 n^3}(p^u)^2 \Phi^{ik}\Phi^{kj} \sin^2(n\tau)\right]\ .}
This sum is divergent, due to the first term.  This is exactly the same as the infinite spreading of a string in a purely flat background\refs{\KKS}, in the unphysical limit where the string is probed at  infinitesimally short length or time scales.
As suggested there, this divergence can be regulated by introducing a finite resolution parameter.  One such definition is to smear $\hat X$ over a small range of $\sigma$,
\eqn\smoper{ \hat{X}^{i}_{\e}(\tau, \sigma) = { 1 \over \e }
\int_{-\e/2}^{\e/2} d \sigma' \hat{X}^{i} ( \tau, \sigma+\sigma'). }
From the expansion \modeexout\ we find the
oscillator expansion
\eqn\modeexout{ \hat{X}_{\e}^{i} =  {i\over 2}
\sum_{ n \neq 0} {1\over n} \left[{2 \sin(n\e/2)\over n\e}
\right] \left[ \alpha_{n}^{\rm{out} \ i} e^{-in(\tau+\sigma)} +
\tilde{\alpha}_{n}^{\rm{out} \ i} e^{-in(\tau-\sigma)}\right]\ .
 }
One finds an equivalent expression if one instead integrates over time, so that $\e$ can alternately 
play the role of a resolution time.
For the oscillators of mode number $n$ with $n\e \ll 1$, the Fourier
coefficients of $\hat{X}_{\e}^{i}$ are essentially the same as those of $\hat{X}$.
But for the higher modes $n\e \roughly> 1$, the effect of smearing
is a suppression of the Fourier coefficient by a factor of $ {N
/n}$, where $ N \sim { 1/ \e }$.   Thus, in practice this regulator plays the same role as the mode cutoff used in \KKS.
As a result, the two point
function in  \deltares\ is regulated:
\eqn\deltaresr{\eqalign{ \ \ \ \invac\Xhat_{\e}^i(\tau,\sigma)
\Xhat_{\e}^j(\tau,\sigma) \vacin = &\sum_{n=1}^\infty
 { {4 \sin^{2} ( {n\e \over 2})} \over {n^{2} \e^{2}} } \cr 
   &\left\{{\delta^{ij}\over 2n} + {1\over4 n^2} p^u\Phi^{ij}
\sin(2n\tau ) +{1\over8 n^3}(p^u)^2
\Phi^{ik}\Phi^{kj} \sin^2(n\tau  )\right\}\ .}}
 The first $N$ terms of the sum in
\deltares\ remain essentially unchanged while the later terms are suppressed
by a factor of order ${N^{2}/ n^{2}}$. Thus the first sum  is no
longer divergent; for  large $N\sim1/\e$ it tends to $|\log \e|$.
It is also possible to estimate the behavior of the second and
third sums. For times $\tau \ll \e$, expanding the sine for small
argument in the first N terms, one finds that the leading behavior
of the second term is $\tau|\log \e |$ while the leading behavior of
the third term is $\tau^2 |\log \e| $.

For $\tau\roughly>\e$, the resolution parameter can effectively be ignored in the second and third terms of the sum; these are convergent, and produce 
 polylogarithms.\foot{The finite sums corresponding to regulated expressions can be written in terms of polylogarithms, Lerch transcendents,  and the polygamma function.}  Specifically, the coefficients of the linear and quadratic terms in $p^u$ are
\eqn\polyone{f_1(\tau) =\sum_{n=1}^\infty {1\over4 n^2}\sin(2n\tau) = {i\over 8} \left[{\rm Li}_2(e^{-2i\tau}) - {\rm Li}_2(e^{2i\tau})\right]}
and 
\eqn\polytwo{f_2(\tau) = \sum_{n=1}^\infty {1\over8 n^3} \sin^2(n\tau) = {1\over 32} \left[2\zeta(3)-{\rm Li}_3(e^{-2i\tau}) - {\rm Li}_3(e^{2i\tau})\right]\ .
}
At short times $\tau\ll1$ these behave as
\eqn\asybeh{f_1(\tau)\approx -\hf \tau\ln\tau + {\cal O}(\tau)\ ,\ f_2(\tau)\approx -{1\over 8} \tau^2 \ln\tau +{\cal O}(\tau^2)\ ;}
at times $\tau\gg1$ they oscillate with amplitude ${\cal O}(1)$.

At large times the stretching parametrized by \deltaresr\ becomes large and our approximations fail.  We therefore focus on the short-time behavior.  Combining our results, we  thus find an asymptotic expression
\eqn\deltaas{\eqalign{\invac\Xhat_{\e}^i(\tau,\sigma)
&\Xhat_{\e}^j(\tau,\sigma) \vacin\cr &\approx \hf \delta^{ij} |\ln\e| + \hf \Phi^{ij} p^u \tau|\ln\e| + {1\over 8}\Phi^{ik}\Phi^{kj} (p^u\tau)^2 |\ln \e|\quad ,\quad\tau\ll\e\cr 
&\approx \hf \delta^{ij} |\ln\e| + \hf \Phi^{ij} p^u \tau|\ln\tau| + {1\over 8}\Phi^{ik}\Phi^{kj} (p^u\tau)^2 |\ln \tau|\quad 
,\quad \tau\gg\e\ .}}

A useful comparison is to a collection of particles that  simultaneously hit the shock wave.  Consider such a collection with label $\alpha$, with some transverse position distribution, and expand about the center of mass of the distribution.  The $\alpha$th particle evolves as
\eqn\alphap{x_\alpha^i(\tau) = x_{cm}^i(\tau) + \delta x^i_\alpha(0) + \hf p^u \tau \Phi^i_j(x_{cm}) \delta x^j_\alpha(0)}
where $\delta x_\alpha^i$ represents the deviation from the center of mass position.  Notice that this follows from the point-particle limit of the string equation of motion, \qeom.
One may compute the average-squared deviation, over the distribution.  Suppose that at impact the particles are distributed such that
\eqn\initdist{\langle \delta x^i(0)\delta x^j(0) \rangle = \ell^2 \delta^{ij}\ .}
Then the subsequent distribution is
\eqn\avsq{\langle \delta x^i(\tau)\delta x^j(\tau) \rangle = \ell^2\left[ \delta^{ij} + p^u \tau \Phi^{ij} + {1\over 4} (p^u)^2 \tau^2 \Phi^{ik}\Phi^{jk}\right]\ .}
Note that in the region $\tau\roughly>\e$, $|\ln\tau| \roughly< |\ln \e|$.  Thus comparing \deltaas\ with \avsq, we find that for a given resolution time the string distribution is bounded within a particle distribution with 
$\ell=\sqrt{|\ln \e|/2}$.

One can perform other checks on this picture.  For example, one can explicitly work out higher-point functions, such as three and four-point functions of the operators ${\hat X}_\e$, and show that they are bounded by the same behavior.  In doing so, one can also include subleading corrections in the $\alpha'$ expansion \phiexp, and finds that they indeed contribute terms suppressed by $1/b$.  Moreover, one can also work out correlators such as $\langle {\hat X}_\e(\sigma, \tau)  {\hat X}_\e (\sigma', \tau')\rangle$ and show that their real and imaginary parts are bounded by the largest of $\langle {\hat X}_\e(\sigma, \tau)  {\hat X}_\e (\sigma, \tau)\rangle$, $\langle {\hat X}_\e(\sigma, \tau')  {\hat X}_\e (\sigma, \tau')\rangle$.

In addition to transverse string spreading, one can also calculate the effect of string spreading in the longitudinal direction.  In flat space, such spreading has been discussed in \SusskindAA.  It can be evaluated via $ \langle  \hat{X}^{v}(\sigma,\tau)
\hat{X}^{v}(\sigma,\tau)  \rangle $, where $\hat{X}^{v} = X^{v} - x^{v}(\tau)$. 
In light cone quantization, $X^{v}$ is determined in terms of the other
string coordinates
\eqn\lcxm{ \hat{X}^{v} = { i \over p^{u} } \sum_{n \neq 0}
\bigg{[}  { L_{n}  \over n } e^{-in(\tau + \sigma)}   + {
\tilde{L}_{n}  \over n } e^{-in(\tau - \sigma)}  \bigg{]}, }
where $L_{n}$ and $\tilde{L}_{n}$ are the light-cone Virasoro
generators. After using the standard expressions for these generators one finds
\eqn\lth{ \langle0| ({\hat X}^v)^2|0\rangle ={ (D-2) \over 6 (p^{u})^{2} }
\sum_{n=1}^{\infty} \big{(} n - { 1 \over n } \big{)} \ .}
Regulating this expression by smearing the fields or equivalently by the mode cutoff $N=1/\epsilon$, one finds
\eqn\ltht{ \langle0| ({\hat X}^v)^2|0\rangle \sim { 1 \over ( \epsilon p^{u})^{2} }\ . }
In the AS background, one can evaluate the quantity
analogous to the flat space case,
\eqn\asprd{ \invac \hat{X}_{\rm{out}}^{v}(\sigma,\tau)
\hat{X}_{\rm{out}}^{v}(\sigma,\tau) \vacin\ . }
Although the algebra is straightforward, one generates a large number of new
terms beyond the flat contribution \lth. One can check that none of the additional terms
provide a contribution larger than the flat space term 
\ltht\ .

\subsec{Black hole formation}

To address whether a trapped surface forms, we ask whether for small enough $\tau$ the string is within the trapped surface, as measured by the squared-deviation \deltaas.  In particular, without the logarithms in \deltaas, the typical string spread is of order 
\eqn\sprest{\Delta X \sim  { G_DE^2\over b^{D-2}} \tau\ .}
This needs to be small as compared to $b$ for the sigma-model expansion \phiexp\ to be valid.  Take $b\roughly< R_S(E)\propto E^{1\over D-3}$ so that the impact parameter is in the black hole region.  $\Delta X$ also has to be small compared to $b$ so that the string is inside the black hole.  This can be achieved by taking 
\eqn\taubda{\tau\roughly<R_S^2/E \sim E^{5-D\over D-3}\ .}
 If we include the logarithmic spreading, a sufficient condition becomes
\eqn\taubd{\tau\roughly<{R_S^2\over E |\ln\e|}\ }
and is still easily satisfied.  Notice that the additive contribution $|\ln\e|$ to the spread is only logarithmically large in $E$ for $\e\sim 1/E^p$, and so does not compete with the power-law spread from the tidal terms in \deltaas.  The logarithmic transverse spread also apparently matches the arguments of \refs{\VenezianoDF,\KaNa} that the gravitational source effectively behaves like a beam of transverse size $\sim\sqrt{\ln E}$.

One can also estimate the effect of longitudinal spreading.  Given \ltht, the string spreading range only exceeds the string length when $\epsilon<1/p^u$.  Thus for a cutoff of size \taubda, the effect is small.

These estimates indicate that in our approximation, and for sufficiently short times, string effects do not distort the size of the strings to scales competitive with the size of the trapped surface.  Thus there is a controlled sense in which each string is, at early times, inside the apparent horizon.  Moreover, notice that for any time $\tau\roughly>\e$, the spread in the string is smaller than the spread in a distribution of particles that starts with a characteristic radius $\ell = \sqrt{|\ln\e|/2}$.  

Thus,  the only apparent way to conclude that a meaningful realization of a black hole doesn't form is if, due to significant modifications of causality, strings can escape from the interior of an event horizon.  This is not strictly ruled out, as causality in string theory is incompletely understood, but is not expected.  Note also that this picture seems to  contradict the suggestion of \refs{\VenezianoER}, that the diffractive excitation component (tidally excited strings) can carry information and energy to future infinity.

\newsec{Discussion}

To summarize the picture:  for impact parameters $b\roughly > (G_D E^2)^{1/D-4}$, scattering is well-described in the Born approximation.  At smaller impact parameters, one instead expects the  eikonal sum \eikamp\ to give a good approximation.  This begins to receive important corrections when strings become diffractively or tidally excited, at $ b_D\sim E^{2/D-2}$.  We expect a reasonable description of the resulting amplitudes could be given by the diffraction-corrected S-matrix discussed in \VenezianoER, which accounts for tidal string excitation, until the regime $b\sim R_S(E)$.

The results of the preceding two sections are consistent with formation of black holes in high-energy string collisions at impact parameters $b\roughly< R_S(E)$, and further support this conclusion by providing a physical description of relevant string and gravitational effects.  In particular, no grounds have been found for string effects interfering with formation of a closed trapped surface, in the ultrahigh energy regime where the Schwarzschild radius significantly exceeds the string length.

Black hole formation indeed appears to be an effect fully governed by {\it gravitational} dynamics.  In particular, it has been argued to be associated with the breakdown of the gravitational loop expansion\refs{\AmatiXE,\AmatiTB}.  In the ultrahigh-energy regime, leading corrections to the eikonal series \eikamp\ arise from exchange of graviton tree diagrams between the high-energy strings.   (There are additional corrections at higher order in $\hbar$.) This produces a power series in $R_S/b$, which becomes divergent at an impact parameter $b\sim R_S$. 

Of course, a very interesting question is what nonperturbative dynamics enters and yields the expected unitary evolution.  The lack of a fully non-perturbative description of the process means that string theory does not yet supply an unambiguous answer to this question.  

It is interesting to consider features of the expected amplitudes.  One is the total scattering cross-section, which is expected to be of the form 
\eqn\sigmaT{\sigma_T\sim  \left[R_S(E)\right]^{D-2}\sim E^{D-2\over D-3}\ .}
Others are  exclusive scattering amplitudes.\foot{For a related discussion, see \ArkaniHamedKY.}  Gross features of a black hole are expected to follow from black-hole thermodynamics.  In particular, given that a particular $n$-particle final state is one of an expected approximate thermal ensemble of $exp\{S_{BH}\}$ states, where $S_{BH}$ is the Bekenstein-Hawking entropy, and $T\sim 1/R_S$ is the temperature, one expects exclusive amplitudes of size
\eqn\elast{{\cal A}(s,t)\sim e^{-S_{BH}}\sim e^{-ER_S(E)}\sim e^{-E^{(D-2)/(D-3)}}\ .}
Refs.~\refs{\FPVV,\VenezianoER} have suggested that such a result might be understood from a divergent time delay associated with the breakdown of the perturbation series.  In particular, as in scattering from a classical black hole, they argue for a time delay of the form $R_S\ln(b-R_S)$.  Below the critical impact parameter, this expression receives an imaginary contribution $\sim i\pi R_S$, which produces an amplitude of the form \elast.  Such a picture also appears to link the result \elast\ with the gravitational nonperturbative dynamics.  A challenge for a nonperturbative formulation of string theory is to produce such amplitudes in finer detail.

The amplitudes \sigmaT\ and \elast\ violate expected bounds for local field theory.   The 
 $D$-dimensional Froissart\refs{\Froiss} bound states\refs{\ChFi\CFV-\PetrovHB} that at $E\rightarrow\infty$,
 \eqn\froissbd{ \sigma_T\leq c (\ln E)^{D-2}\ ,}
 with constant $c$, and expected bounds of the
 Cerulus-Martin\refs{\CeMa,\Mart} form constrain fixed angle asymptotics,
\eqn\CMbd{|{\cal A}(s,t)| \geq e^{-f(\theta) E \ln E}\ ,}
for some function $f(\theta)$.
 While  basic assumptions needed to derive these bounds, particularly the existence of a gap, are not strictly satisfied in gravitational scattering, it it tempting to conclude that violation of such expected bounds is associated with some essential nonlocality associated with non-perturbative gravitational dynamics\refs{\LQGST,\GiddingsBE}.

\bigskip\bigskip\centerline{{\bf Acknowledgments}}\nobreak

The authors wish to thank  N. Arkani-Hamed, H. Elvang, R. Sugar, and G. Veneziano  for discussions.
The work of SBG and AM  was supported in part by the Department of Energy under Contract DE-FG02-91ER40618,  and by grant  RFPI-06-18 from the Foundational Questions Institute (fqxi.org).  The work of DJG was supported in part by the National Science Foundation  under Grant No.  PHY99-07949.

\listrefs

\end